# Smart Policies for Artificial Intelligence

Miles Brundage (miles.brundage@asu.edu) and Joanna Bryson (Bryson@conjugateprior.org)

Over the past few years, developments in artificial intelligence (AI) have captured the imagination of tens of millions, if not billions, of people around the world. You will have seen countless news stories about the proliferation of driverless cars and AI systems beating the world champions at Jeopardy! and Go – not to mention science fiction stories about more advanced, if often implausible, AI systems. But AI is not just in the news.  You use AI when you use a search engine, use GPS for navigation, or use the voice recognition feature on your smartphone.  You personally, probably every day, interact with technologies made possible only by scientific research in AI.  Yet despite the explosion of pervasive AI technology, and the often overstated hype and hysteria about AI in the media, there's been too little substantive discussion of one of the most important question about our present and future with these technologies: what can and should policy makers do to ensure that we reap maximum benefits – and avoid hazardous pitfalls – of AI?

To some, the idea of policy specifically oriented toward AI seems premature, misguided, or even dangerous. For example Andrew McAfee, co-author of the best-selling book *The Second Machine Age,* wrote "it's way too early for explicit AI policy."[i] These remarks were echoed by Demis Hassabis, CEO of Google DeepMind, the developers of AlphaGo – the system that defeated Lee Sedol 4-1 in a Go match in South Korea. In an interview with BBC, he said, "I think it's much too early to think about regulation."[ii] Two points are often made in defense of postponing serious AI policy analysis and implementation. One is that AI is still in the early stages; the other is that misguided policy (or, as it's often derogatorily reduced to, "regulation") could hamper the development of the technology and thereby forestall the substantial societal benefits of AI.

Both these arguments are flawed.  First, AI is already sufficiently mature technologically to impact billions of lives trillions of times a day.  Far from being a fiction or laboratory curiosity of interest only to scientists and engineers, AI already undergirds major segments of the economy.  Alphabet – parent of Google and the company widely considered to be at the forefront of AI research – is the largest company in the world in terms of market capitalization, and got there by using AI technology to deliver search results and advertisements to billions of Internet users.  Google CEO Sundar Pichai now envisions an "AI-first world," where natural human speech and gestures will replace mobile phones and tablets as the primary interface to technology.[iii] Moreover, the so-called "sharing economy" relies on smart interfaces; Uber and Lyft in particular rely on GPS to give ordinary people the capacity to at least navigate like a professional driver. Those routes, in turn, are determined using algorithms derived from AI research.  Apple uses advanced AI to make the extraordinarily complex robotic device that is an iPhone usable by millions of people, with speech recognition algorithms being just one example.

Second, science and technology policy operates much more broadly than restrictive regulation, and even regulation can be pro-innovation (such as by developing and enforcing standards that enable cross-industry collaborations). Policy encompasses a wide spectrum of possible interventions to accelerate, decelerate, or simply change the developmental trajectory and diffusion of a given science or technology field. Policy does not require the full flourishing of a technology before it becomes important, and indeed it can be most important in the earlier stages of development. Policy simply refers to a set of decisions that societies, through their governments, make about what they do and do not want to permit, and what they do or do not want to encourage. In the case of AI, a number of such decisions are already being made, and many more will be made in the future. We will be best equipped to do so after realizing the breadth and depth of *de facto* AI policy that already exists, and after appraising the unique (and not so unique) new challenges that AI raises for society, most of which have not yet been addressed by policy-makers.  These are the objectives of this article.

Both the sophistication and the societal impact of intelligent technologies are set to increase substantially in the coming years and decades, as are the associated policy challenges. To give just a taste of the policy issues raised by AI, consider some of the following questions:  how is liability to be allocated across complex hardware/software supply chains involved in developing driverless cars, and between those designers and the end users who may abuse these features or misunderstand their systems' capabilities? What can be done to forestall the looming (if not ongoing) arms race between countries to develop lethal autonomous weapons? How can government agencies protect consumers and citizens from unethical, unsafe or unsound use of AI systems employed in critical contexts such as health, finance, or employment by companies or individuals.  How do we avoid negligent or intentional bias or harm caused by individuals or groups with access to systems with personalised data or involved in individual decision making? How can the economic disruption and creation associated with AI's workforce impact be constructively managed? And more broadly, how can governments ensure that AI will benefit the many, rather than the few?

While this list may seem daunting and will likely involve a wide range of decisions in and outside government over the coming years, there are many things that can be done today to improve our ability to respond appropriately to these developments. Encouragingly, in America the White House Office of Science and Technology Policy is currently conducting a series of public workshops on questions of AI and policy, following on from similar exploration and public dissemination which has been ongoing in the European and British Parliaments. We have written this article hoping to inform next steps by governments in formalizing, integrating, and improving AI policy, as well as to inform business leaders and ordinary citizens.

As we mentioned, there already exists what we call *de facto AI policy* – a patchwork of policies impacting the field's development in myriad ways, though only a few of which are already explicitly oriented toward AI. **The key question is not whether AI will be governed, but how it is currently being governed, and how that governance might become more informed, integrated, effective, and anticipatory.**  As a society, we already know enough about some critical issues, and we could take proactive steps to learn more. As authors, we make recommendations concerning these below.  But first, we start by explaining what AI is and reviewing the key components of its de facto policy regime. Then we show how lessons from the governance from other technologies, and early lessons from de facto AI policy, can inform a smart approach to AI policy that continues to foster innovation while safeguarding society as citizens and consumers.

**The Nature of AI and Why it Matters**

Throughout its history, AI has been defined in various ways, as have terms like "robots". One might come away from journalistic accounts of AI with the impression that it is either meaningless or an aspiration for future developments that may never arrive. In fact, AI is a mature scientific discipline with myriad existing and future applications, and ambiguities about its precise definition have hardly impeded its progress.  Nevertheless, ambiguities in definitions do impede communication, so we will set out a precise set of usages at least for the purpose of this article.

*Intelligence* is simply the capacity to express an appropriate behavior in response to changes and opportunities in one's environment.  Anything that creates change is technically termed an *agent*, even chemical catalysts. Agents, whether animal or artifact, are considered more intelligent when they can more successfully achieve more goals in more varied environments. Intelligence therefore admits of degrees, and also multiple dimensions e.g. semantic knowledge vs dexterity.  Further, intelligence can be divided up into stages in the decision-making process, such as perception, reasoning, and action.  Artifacts that implement any of these parts are said to be AI, and can augment or substitute for humans in any or all of these capacities in particular contexts.  Organizations, too, can be considered intelligent entities in that (through their constituent parts) they perceive their environments, process this information, and then act.[iv]

Many AI-based systems are already in wide use.  Search engines aid our perception – and increasingly our reasoning – by processing large amounts of data and presenting an otherwise chaotic Internet in the form of human-interpretable search results. Driverless cars perceive their environment through sensors, process this information (think), and then act through steering, acceleration, and braking. Map routing algorithms underlying modern GPS-based navigation systems derive from AI research on the problem of search, particularly the A* ("A star") algorithm which also underlies the navigation skills of characters in both computer games and some heavily CGI films. A* has been around as an algorithm for decades, although improvements continue to be made, what's changed most is the availability of compute power, and also the data about the real world for the AI to search over.

Much recent excitement about AI has focused on "deep" multi-layer machine learning approaches, typically using artificial neural networks, though also reinforcement learning – again, algorithms and representations that have been both researched and implemented in products for decades.  Machine learning extracts useful patterns from data.  Depending on what the data represents, these patterns can improve any or all of perception, thinking, and acting.  In brains as well as modern AI, perception and action are often tightly coupled: for example, DeepMind, years before the Lee/AlphaGo match, came to the world's attention for developing a system that could learn to play a wide variety of Atari videogames as well as a human using only the image on the screen and the score as inputs.

Artificial intelligence is not necessarily similar or equivalent to human intelligence.  In fact, because human intelligence keeps evolving (primarily culturally but even biologically) to meet the requirements of our animal lives and societies, it is unlikely that even if an AI was built to be exactly like human intelligence that it would stay that way for long. Natural intelligence consists of many strongly-coupled computational components with specialised biological implementations, linked with specially-evolved and then individually-developed sensors and actuators both fitting and defining the ecological niche of the species that evolved them. AI systems tend to be restricted to fewer computational representations, which are kept relatively isolated from each other to make them cheaper to manufacture and easier to debug. While sharing neither the complex grabbag of core organising motivations of humans nor our particular suite of sensors and effectors or capacities to self-heal, artificial intelligence is not strictly less than human intelligence. AI is able to master tasks that would be inconceivable for individual humans, for example indexing the entire web, or translating between hundreds of languages. **AI is not just fiction, nor is it a newly evolved life form meeting its own spontaneously-generated goals. It is a well-understood body of research and practice** that gradually improves our (and our institutions') abilities to perceive, think about, and act upon the world. Improvements in computer hardware, the availability of large datasets, and algorithmic improvements, along with substantial corporate investment, have all accelerated progress radically in recent years, and there are few if any signs that AI progress will decelerate or stall soon.

AI is importantly like many existing technologies. For example, like computing technologies more generally, AI relies on both software and hardware developments, and is widely applicable in a range of contexts. Like electricity (an analogy we return to later), it is a general purpose technology in the economists' sense, one that can improve productivity in a great many, if not all, industries. Like many technologies too, AI relies for its implementation on considerable expertise, though the barriers to entry has steadily declined in recent years as increasingly user-friendly tools have become available. However, AI has several differences from most prior technologies that make it especially difficult but also important to govern wisely. AI allows the automation or augmentation of human activities that have previously been beyond the reach of technology, extending computing's reach beyond the structured, predictable environments and into a wider range of human domains. Also, the very nature of intelligence, especially at high levels, is that it takes action, and in a way that is to some extent independent and unpredictable in terms of the precise actions that will be taken by a system at a particular time. This is part of the basis for AI's appeal, in that we can use it to solve problems we couldn't otherwise solve, or just reduce the amount of our attention taken by easily solvable tasks, but it also raises the possibility of an unexpected behavior proving dangerous for reasons that were not foreseen. Moreover, as emphasized by many scholars, the social nature of some AI-based robotics (and even merely software) technologies creates the possibility for human attachment in a way that other technologies do not.[v] These characteristics make it particularly important that governing bodies are aware of the state of the art of the technology and prepared to prevent its abuses.

**De Facto AI Policy**

While AI research conducted by employees of some of the major technology companies (Alphabet, Microsoft, IBM, Apple, and Amazon, to name a few) has gotten much of the attention in recent years, this work builds on and continues to benefit from decades of government investment in AI. Indeed, the initial Dartmouth conference where the term "artificial intelligence" was coined was funded by a US government grant, as was the Google founders' graduate research and the research that led to the development of Siri. Currently, in response to the development of driverless cars, federal and state governments now have implemented various legal policies that pertain to forms of AI-based technologies. Similarly, while drones need not involve AI, some of them do, and there is now a comprehensive set of federal policies governing drone usage. These examples of both funding for research and explicit regulation of AI-related technologies at the federal level are part of the patchwork of de facto AI policy currently in place. In this section, we'll summarize some of the key components of de facto AI policy, before discussing how it can be improved.

Roughly, de facto AI policy falls into three categories. First, there are policies that are specifically oriented toward governing AI-based technologies, such as driverless car regulations. We call these *direct AI policy*. Second, there are policies that indirectly affect AI-based technology development, but are nominally focused on other technologies or technology in general, such as intellectual property laws—we call this indirect AI policy. Third, there are policy domains in which AI development is neither specifically targeted nor significantly affected, but in which knowledge of plausible AI futures would benefit policy-makers, such as education, urban planning, and welfare policies—we call these AI-relevant policy. In all of these areas, there is room for improvement in terms of the competence of government agencies, the degree of foresight applied to decisions in these domains, the extent to which they are open to democratic scrutiny and influence, and the extent to which decisions are made in an integrated fashion. Explicit AI policy takes several forms. Perhaps most significantly, the U.S. federal government is a major funder of AI research and development. Much of this funding comes from defense and intelligence agencies such as DARPA, IARPA and the offices of scientific research of the various branches of the military, but substantial amounts also come from the National Science Foundation and, increasingly, other civilian agencies. Much of this funding is fairly general purpose, in that it encourages development of AI methods–and indeed AI talent–that can be applied to nearly any industry or domain of life. There is also a good deal of government support for targeted AI development, focusing on specific capabilities of interest to policy-makers, such as for AI-based surveillance and improving (or fully automating) military decision-making. Much recent development in AI and robotics is traceable directly to such federal investments – the DARPA-sponsored driverless car competition helped advance the state of the art and usher in the large-scale corporate investment we see today. A similar DARPA competition occurred recently, this time for humanoid robots operating in in human-built environments, and lessons learned from that competition will catalyze research and investment for years to come (indeed, a winner of one round of the competition was bought by Google).

Beyond federal support for AI research, there are also many legal policies that specifically relate to AI-based technologies. For example, several states as well as the federal government now have policies pertaining to driverless cars—the [Department of Transportation], for example, responded with an affirmative to a request from Alphabet for clarification on whether its autonomous cars can themselves be considered drivers for legal purposes.[vi] States, in turn, have developed a patchwork of laws governing specific aspects of driverless car usage such as their testing and the role of humans in (or out of) the loop in decision-making.[vii] Moreover, the Department of Defense, in part spurred by a vibrant movement in civil society for a ban on lethal autonomous weapons, has clarified its stance on such weapons at different points, although ambiguities remain and the question of such a ban remains a hot issue. Modern AI tends to draw heavily on large datasets, and regulations concerning the privacy and use of personal data by corporations and governments are thus often also AI policies. As a final example, as alluded to previously, drones are often human-operated by remote, but since they can often can make some decisions without human intervention, federal, state, and local drone regulations can also often be considered direct AI policies. Notably, although our examples here focus on direct AI policies in the U.S., efforts to govern AI more thoughtfully are ongoing in many countries around the world, especially in the European Union but also South Korea, Japan, and elsewhere.

A much wider range of policies that we call *indirect AI policies* impact AI development and diffusion, even though they are not explicitly oriented towards AI. For example, intellectual property laws enable large and small technology companies and individuals to patent AI algorithms and applications, affecting which commercial products and services are likely to be developed and who will or won't be able to access them. STEM education policies, especially at the high school and college levels, critically affect the supply of talent into the AI field. The inability of STEM education in general and computer science education in particular to keep up with the rising demand for AI talent has led to a very high premium for new graduates of computer science programs, and has opened an opportunity for a wide variety of online education courses and "nanodegrees" in AI areas such as machine learning. Additionally, labor policies related to health insurance portability and independent contractor status affect the extent to which companies involved in the "sharing economy" are able to thrive. Other labor policies like minimum wage laws also affect the incentive of companies to invest heavily in automation, by changing the relative economic returns for labor versus capital expenditures (which includes AI-based robotics systems). Tax credits for research and development affect industrial investment in AI. Liability laws affect the extent to which companies are liable for AI-related malfunctions and may determine in which states and countries we see AI developed and trialed, and the robustness of our consumer protection agencies influences whether we are likely to see corporations held accountable for deceiving customers about the capabilities of AI-based devices (a major issue we will return to below when discussing our recommendations).

Finally, *AI-relevant policies* are those decisions governments make that do not immediately or directly affect AI development or diffusion, but which still have sufficient impact that they should be informed by it. Two examples here are illustrative – general education and health policies. First, education policy is not only an indirect AI policy in that it affects the availability of AI developers, but it is also an area that needs to be informed by our expectations about plausible AI futures. There is a growing consensus that, while AI may not overall reduce the number of jobs available in the economy, it will cause substantial disruption for particular occupations, regions, and companies, and will improve the fortunes of others. AI technology is advancing at a rapid pace and will be considerably more advanced in 15-20 years, yet education policies that operate over that same timeframe are largely still being implemented as if nothing is going to change. While there are exceptions, such as government investments in computer science education motivated by a vision of a growing role for computing in the workforce, these are mostly motivated by the more obvious indirect AI policy concerns. These are sensible, but stop short of the sort of deeply rethinking which education topics and competencies to emphasize or deemphasize when preparing students for a world that may be very different. Second, health policies are still playing catch-up with regard to electronic medical records, and have not even begun to reckon with the enormous potential of AI and robotics to improve outcomes, as well as to decrease costs through automation (or to increase it, as standards rise). As with education, one can find isolated examples of innovation in this area, such as National Institutes of Health investment in some robotics projects, but they are drops in the pond of overall health care spending, when what is needed is a broad vision of how to learn from the failures of the last wave of digitization in health care while preparing for the next one. Similarly, although some cities are considering the consequences of driverless technology on the costs of public transportation (which are currently dominated by staffing) urban and suburban planning have not really begun to consider the full consequences of the AI revolution, and how citizens and communities may respond to the possibilities it affords, particularly for achieving more sustainable and lower-cost living.

These are just a few examples of areas where a forward-thinking perspective informed by the latest (and foreseeable future) advances in AI could pay dividends in policy robustness, effectiveness, and cost savings. Other examples could surely be given – for example, there are substantial opportunities for AI to improve the monitoring and repairs of critical infrastructure, to accelerate research into clean energy technologies, and to generally improve the efficiency of government operations. Yet we are unlikely to see such rigorous planning in the near future if steps are not taken to bolster government capacity for AI governance and make other key changes and investments. It's to these steps that we now turn.

**Learning from experience**

While machines are increasingly capable of learning from experience, governments at times seem to begin governing new technologies anew without learning from other relevant cases. And despite the clear signs that AI is developing quickly, diffusing rapidly, and likely to continue having substantial societal impacts well into the future, there has yet been no substantial effort to improve U.S. government capacity to appropriately govern AI. As of the time of our writing, the White House's exploration into the issue remains in an information-gathering stage–we hope they and other policy makers will consider some of the recommendations below when deciding on bolder next steps. Our recommendations are organized around several core areas in which de facto AI policy is currently lacking.

First, there is the question of **government expertise** in AI. The most important and obvious thing that governments should do to increase their capacity to sensibly govern AI is to improve their expertise on the matter through policy changes that will support more talent and the formation of an agency to coordinate AI policy. It is widely recognized that governments are ill-equipped currently to deal with ongoing AI developments.[viii] A telling example of this deficit is that NASA had to be brought in to aid the Department of Transportation in investigating reports of unintended acceleration due to Toyota cars' control systems.[ix] This incapacity has two forms: dispersion of expertise across many agencies, and the low level of overall expertise even when adding up all those people. To some extent, dispersion is inevitable and desirable – many agencies deal with AI-relevant policy issues, and it would be undesirable to concentrate all of the experts in one place as a result. However, compelling arguments have been made by legal scholar Ryan Calo that a Federal Robotics Commission would be appropriate for coordinating governance of robotics and accelerating the accumulation of federal expertise in robotics.[x] We support something along such lines for AI more generally, given the close connections between AI and robotics, perhaps a Federal AI and Robotics Commission would be an appropriate title for an agency addressing both areas and their many intersections. The long history of governance of emerging technologies suggests that new agencies can play a constructive role in partially centralizing expertise in a topic, improving coordination, and clarifying the allocation of responsibilities. Many important questions have been raised about how such an agency should be designed, where in the government it should sit, and what its responsibilities should be.[xi] The other aspect of government's lack of expertise, the total amount of expertise in government as opposed to its dispersion or concentration, is perhaps more challenging to address. But here, too, there are useful precedents. The challenge is partly one of incentives – AI is a very hot commercial and academic area right now, and salaries available in industry greatly exceed those in government. Moreover, while some agencies have recently made great strides on the closely-related issue of the prestige of government jobs (through e.g. dynamic and mission-oriented agencies like the U.S. Digital Service), most of the top students in AI still presently go to academic and industry jobs rather than choosing to serve in government. This can be addressed in part by learning from the experience of other agencies, such as the Advanced Research Projects Agency – Energy (ARPA-E), the legislation for which specifically enables it to hire top talent rapidly and at competitive salaries. Thus, while there are steps that can be taken by the executive branch today to improve capacity in AI, tailored legislation may be needed to scale up government AI expertise by bypassing byzantine hiring policies that deter people from considering government jobs when easier-to-obtain and higher-paying jobs are available elsewhere. And while pursuing direct employment of AI experts by the government is important, developing innovative ways of tapping into the outside expertise of industry and academia can be pursued in parallel, in recognition of the (partially) inherent difficulties outlined above.

      The second key area for improvement in AI policy is in the **funding of research**. While there is much to praise about current AI funding policies in the U.S., such as their critical role in supporting basic research and graduate student training, there are several ways in which funding policies could be improved. First, the U.S. should substantially increase the share of civilian AI research relative to defense and intelligence AI research. The latter, while often useful for the reasons outlined earlier, is less likely to address significant societal problems in the areas like health and energy than funding specifically targeted at such problems.  Of course, there is significant private funding available for AI research, but most of this is provided either directly or indirectly in a commercial setting.  Particularly for research concerning the possible impacts of commercial AI on society, more neutral funding directly accountable to voters may be of use. Many AI researchers in the United States pursue pro-social research on their own time or on the few grants available for such purposes, but many others (sometimes grudgingly) rely on military and intelligence funding. Besides providing additional options to researchers, additional civilian funding would provide an opportunity for concerted efforts–coordinated across agencies–to tackle societal problems in part through the development and diffusion of AI. Second, research agencies should dedicate funding to research the ethical, policy, and safety issues raised by artificial intelligence. To date, much of such research is done by researchers on their own time, or through private investments.  Again, this is a clear area where government should support the production of public goods, such as insights into risks and opportunities of AI, means of safely controlling advanced AI systems, research on transparency and understanding of AI systems, and technological forecasting. Additionally, work on the ethics of

AI and the policy issues surrounding it, ranging from privacy to potential security risks, should be more actively supported. While such research would be broadly beneficial, it is not currently being pursued by more than a handful of unusually forward-thinking companies and think tanks, and a few government-funded projects.

Importantly, in funding such research, agencies should learn from the experiences of the Human Genome Project, in which Ethical, Legal, and Social Implications (ELSI) research was widely regarded as not having affected either policy or ongoing technical research. In contrast, AI research in ethics, policy, forecasting, safety, and other areas should, to the extent possible, be integrated into technical research projects, supporting truly needed interdisciplinary research. Research grants should include provisions for, and additional earmarked funding to support, ethical training for undergraduate and graduate student researchers. While some AI researchers are proactive about raising ethical issues in their classrooms and laboratories, and there is effort in AI at the federal level to ensure such training, many students enter industry and research positions with little awareness of the complex ethical dilemmas they may face. Such investments and mandates should include pilot programs to pursue innovative approaches to integrating ethics into AI curricula, and should learn from the experiences of prior efforts in other areas such as biology.

A third area for policy research improvement relates to the **diffusion of AI technologies**. As we can learn from domains like health care and energy technologies, it is not sufficient to fund basic research and expect it to be widely and equitably diffused in society by private actors. In energy, a long history of experience tells us that public-private demonstration projects, tax incentives, and other approaches are critical complements to basic research funding.[xii] In the case of AI, what's notable is the generality of AI techniques. With sufficient expertise, they can be applied to virtually any domain, but if private industry leads diffusion of such technologies, we can expect a deficit in areas where AI may not quickly save large amounts of money, or lead to a sustainable business. Yet such areas may be of critical social value—for example, in areas of poverty alleviation, rare diseases, and accelerating clean energy technologies. AI in this sense is somewhat analogous to electricity, as argued by AI luminary Andrew Ng – it can increase productivity in a wide range of areas, but what often goes unmentioned in this analogy is that it took many decades for electricity to reach some markets, and indeed, over a billion still lack access to it.[xiii]

Currently, we see large amounts of money flowing into AI research and development, but much of that investment is concentrated in a few key domains like driverless cars. The scarcity of expertise makes it conceivable that the majority of the societal potential of AI will go untapped for well into the future. Fortunately, there are encouraging case studies that suggest a different approach to diffusing AI, and they may provide government with leverage to simultaneously make progress on the ethical and safety fronts discussed earlier. Specifically, the U.S. federal government could develop a new agency inspired by agricultural extension services and the Manufacturing Extension Partnership (MEP). A so-called AI Extension Service, or something similar, could fulfil three functions simultaneously. First, it could provide a point of contact for non-profits, local governments, and small businesses that lack AI expertise to seek out assistance with developing AI applications appropriate to their needs, providing a complement to industry that can focus more on unprofitable or especially societally-important areas that are unlikely to be addressed by private forces alone, or require more public trust or long-term stability than any individual corporation can provide. Second, such an agency could aid with the expertise gap outlined earlier, by providing an exciting career path for AI experts, in which–like in industry–they can help people solve real problems, but for a wider range of problems than they might be exposed to when profitability necessarily guides decision-making. Already, the U.S. Digital Service has effectively recruited many top IT experts to public service with a similarly exciting mission, though focused on improving government processes rather than serving non-government entities, and without a strong AI focus. An AI Extension Service could either be an additional growth area for that agency or an inspiration for a more AI focused, outward looking agency. Third, by serving as a repository of expertise, such an agency would have the ability to influence the specific means by which AI is adopted in the marketplace – for example, the ethical and safety standards, the transparency of AI systems, and user and public engagement in developing the systems. Depending on the scale of this initiative, such an effort may or may not affect a large share of AI applications directly, but it could at the very least serve as an exemplar for private projects, and a proving ground for best practices.

The fourth and final area of AI policy we'll discuss involves **planning for the future**. While much is known about the state of the art in AI today and how it might evolve in the near future, there are still substantial uncertainties. No one knows for certain, for example, when and whether AI systems will be able to broadly compete with human intelligence across a wide range of cognitive areas – and yet, this uncertainty does not mean we should ignore such a possibility. Rather, we suggest that a Federal AI and Robotics Commission, or existing institutions like the White House Office of Science and Technology Policy, coordinate an on-going government-wide initiative to investigate possible AI futures, drawing on the best methods in technological forecasting and scenario planning, and to also investigate both how agencies can leverage such technologies for the public good, and how critical mission areas could be affected by AI should adjust based on such futures.  For one example, in the case of the Department of Education, preparing future workers and citizens for coexistence with and enhancement by AI.  The goal here should not be picking one preferred future and seeking to achieve or adapt to it (though, of course, agencies should be proactive) but rather seeking robustness and preparedness in the face of uncertainty.

**Toward smart AI policies**

In this article, we have reviewed the nature of AI and why it is of critical social importance.  We have refuted the claim that it is too early for government policy intervention in AI by showing its extensive existing impact on society, and the broad range of policy already affecting AI's development and dissemination.  and the , Further, we have made specific policy recommendations based on lessons learned (both positive and negative) from previous government support and intervention in important technologies, such as energy and the human genome. Many issues have not been addressed here, such as how and to what extent to democratize AI policy—how can we ensure that citizens can play a role in influencing the development of these systems which already influence them so heavily? And how can we ensure that humans remain accountable for the actions of AI systems, while also fostering continued innovation in areas where autonomously acting systems can have significant societal benefits, such as autonomous cars? These are just some of the questions that remain. But by taking the steps we suggest—building expertise in government, more thoughtfully funding research, creating specific agencies to monitor and encourage social impact, taking a proactive approach to diffusing AI technologies, and generally planning better for the future—the government can provide a foundation for a robust, forward-looking AI policy system. Building on these recommendations, we believe national and trans-national governments can help society to reap the benefits while reducing the downsides of artificial intelligence.

**Recommended reading**

Brynjolfsson, E. and McAfee, A. 2014. *The Second Machine Age: Work, Progress, and Prosperity in a Time of Brilliant Technologies*. New York: W.W. Norton & Company.

Bryson, J. 2016. "Patiency is Not a Virtue: AI and the Design of Ethical Systems," *Proceedings of AAAI 2016.*

Calo, R. 2014. "The Case for a Federal Robotics Commission," Brookings Institution.

Calo, R., Froomkin, A. M., and Kerr, I. eds. 2016. *Robot Law*. Cheltenham: Edward Elgar Publishing.

Lin, P. et al. eds. 2011. *Robot Ethics: The Ethical and Social Implications of Robotics.* Cambridge: The MIT Press.

Parry, V. et al. 2013. "Principles of Robotics: regulating robots in the real world," Engineering and Physical Sciences Research Council (EPSRC), U.K. Available at https://www.epsrc.ac.uk/research/ourportfolio/themes/engineering/activities/principlesofrobotics/ Robot ethics

Walker Smith, B. 2016. "How Governments Can Promote Automated Driving," Available at SSRN: http://ssrn.com/abstract=2749375

**Author biographies**

Miles Brundage is a PhD candidate in Human and Social Dimensions of Science and Technology at Arizona State University.

Joanna Bryson is a Reader (associate professor) in Artificial Intelligence at the University of Bath, and a visiting Fellow at the Princeton Center for Information Technology Policy.

**Acknowledgments**

The authors gratefully acknowledge helpful feedback from Dan Sarewitz, Kevin Finneran, and Seth Baum. Miles Brundage's work on this article was supported by the National Science Foundation under award #1257246 through VIRI. The findings and observations contained in